\documentclass[twocolumn,showpacs,superscriptaddress,prb,floatfix]{revtex4}
\usepackage{amssymb,amsbsy,amsmath,bm,graphicx,wick,ulem}
\usepackage[usenames]{color}
\newcommand{\cF}{\mathcal{F}}

\newcommand{\br}{\mathbf{r}}
\newcommand{\cG}{\mathcal{G}}

\newcommand{\hI}{\hat{I}}
\renewcommand{\Im} {\mathop{\mathrm{Im}}}
\renewcommand{\Re} {\mathop{\mathrm{Re}}}
\begin{document}
\title{Ohmic and non-Ohmic Andreev transport through an interface between 
  superconductor and hopping insulator: Dramatic role of barrier properties}
\author{M. Kirkengen}
\affiliation{Department of Physics, University of Oslo,  P. O. Box 1048
  Blindern, 0316 Oslo, Norway}
\author{J. Bergli}
\affiliation{Department of Physics, University of Oslo,  P. O. Box 1048
  Blindern, 0316 Oslo, Norway}
\author{Y. M.  Galperin}
\affiliation{Department of Physics, University of Oslo,  P. O. Box 1048
  Blindern, 0316 Oslo, Norway}
\affiliation{Center for Advanced Materials and Nanotechnology at the
  University of Oslo, Argonne National Laboratory, 9700 S. Cass Ave.,
  Argonne, IL 60439, USA, and A. F. Ioffe
Physico-Technical  Institute, 194021 St. Petersburg, Russia}
\date{\today}
\begin{abstract} 
At low temperatures and voltages tunneling transport through an
interface between a superconductor and hopping insulator is dominated
by coherent two-electron tunneling between the Cooper-pair condensate
and pairs of localized states, see Kozub \textit {et al.}, \prl
\textbf{96}, 107004 (2006). By detailed analysis of such transport we
show that the interface resistance is extremely sensitive to the
properties of the tunneling barriers, as well as to asymptotic
behavior of the localized states. In particular, dramatic cancellation
takes place for hydrogen-like impurities and ideal barrier. However,
some disorder can lift the cancellations restoring the interface
transport. We also study non-Ohmic behavior of the  interface resistor
and show that it is sensitive to the Coulomb correlation of the
occupation probabilities of the involved localized states. It is
expected that non-Ohmic contribution to $I-V$-curve will experience
pronounced mesoscopic (fingerprint) fluctuations.
\end{abstract}

\maketitle

\section{Introduction}

In this paper, we address the charge transfer through the interface
between a superconductor (SC) and a hopping insulator (HI), i.e., a
system where transport occurs via hops between localized
(\textit{non-propagating}) single particle states. There are many
experimental situations in which the hopping insulator is coupled to a
measuring circuit via superconducting leads, see, e.g.,
Ref.~\onlinecite{Rentch}. However, it has long been known that
transport of single electrons into or out of a superconductor is
exponentially suppressed at low temperatures as $e^{-\Delta/T}$, due
to the energy gap, $\Delta$, of the superconductor.\cite{Giaever}
Consequently, single-electron tunneling cannot be responsible for
the charge transfer between a superconductor and any normal conductor.

The problem of charge transfer between a SC and a HI was first
addressed in Ref.~\onlinecite{Kozub1}, where it was shown that at low
temperatures the transport is governed by the \textit{time-reversal
reflection}, where electrons tunnel into superconductor from the
localized states in the hopping insulator located near the
interface. This process is similar to so-called \textit{crossed}
Andreev charge transfer discussed previously in connection with
various mesoscopic systems.\cite{CA,spin-entangler} Electrons hopping
from the superconductor to impurities near the surface of the
insulator and back again have been proposed as a source of qubit
decoherence for some systems.\cite{Bergli}

In this paper we extend the analysis of Ref.~\onlinecite{Kozub1} in
two directions.  \textit{Firstly}, we consider the influence of the
properties of the tunneling barrier on the charge transfer, which
turns out to be surprisingly sensitive to the barrier
roughness. Namely, we found that interference effects in tunneling can
lead to a significant \textit{increase} in the interface resistance
due to fine cancellations of the contributions to the two-particle
tunneling probability. The roughness of the barrier suppresses these
effects and in this way influences the interface resistance.
\textit{Secondly}, we consider \textit{non-Ohmic} transport through
the interface. We will show that the interface contribution to the
resistance can be strongly nonlinear, and that the nonlinear behavior
is essentially related to the Coulomb correlation of the occupation
numbers of the localized states adjacent to the interface. One can
expect a rich pattern of reproducible (fingerprint) fluctuations in
the $I-V$-curve due to pronounced non-Ohmic contributions of
individual pairs. Thus, combined studies of the  linear interface
resistance and average nonlinear $I-V$ curve and its fluctuations  may
tell a lot about barrier details and about the formation of the
depletion zone near the barrier.

The paper is organized as follows. In sec.~\ref{Model} we set the
stage presenting the model of Ref.~\onlinecite{Kozub1} for the
coherent charge transport. Detailed calculations for the Ohmic case
are presented in Sec. III where we show how the interference-based
cancellations occur.  In Sec.~\ref{linearconductance} we discuss
several generalizations of the model, including the effects of
modification of the tails of localized wave functions and barrier
details. In Sec.~\ref{nonlinearconductance}, starting from calculation
of the non-Ohmic contributions of the individual pairs, we demonstrate
how account of the Coulomb correlations leads to non-Ohmic behavior of
the interface conductance.

\section{Model} \label{Model}

As shown in Ref.~\onlinecite{Kozub1}, the contact resistance can be
governed either by the interface tunneling barrier, or by deformation
of the hopping cluster in the HI in the vicinity of the interface. Here,
for brevity, we will focus on the situation where the interface
resistance is dominated by the barrier.

We start with the case of linear conductance where it is natural to
use the Kubo linear response theory.~\cite{Kubo} According to this
theory, the conductance, $\cG$, is expressed through the
susceptibility,
\begin{equation}\label{in} \chi (\omega)=i \int\limits_{0}^{\infty}
\left\langle \left[  \hI^{\dag}(t),\hI(0)\right]  \right\rangle
e^{i\omega t}\, dt
\end{equation} as $\cG=\lim_{\omega \to 0}\omega^{-1}\Im
\chi(\omega)$. Here $\hI(t)$ is the current operator and we will use units where $\hbar=1$.

Let a superconductor and a hopping insulator occupy adjacent 3D semi-spaces
separated by a tunneling barrier (TB). The presence of the barrier
simplifies calculations which will be made in the lowest non-vanishing
approximation in the tunneling amplitude $T_0$. This models the
Schottky barrier usually present at a semiconductor-metal
interface. Then the current operator is defined
as:~\cite{tunnel-current}
$$ \hI(t)=ie \int d^{2}r\, d^{2}r' T(\br,\br')[ a^{\dag}(\br,t)b(\br',t) - \text{h.c.}]
\, , $$
where $\br$ is the coordinate on the superconductor side of the interface plane,
$\br'$ is the coordinate on the semiconductor side,
$a^{+}(\br,t)$ and $b(\br,t)$ are creation and annihilation
operators in the semiconductor and superconductor, respectively,
$d$ is the electron localization length under the barrier.

The Matsubara formalism, see e.g. Ref.~\onlinecite{AGD}, allows
calculation of the susceptibility by analytical continuation of the
so-called \textit{Matsubara susceptibility} defined as
\[
\chi_M(\Omega)= \int^\beta_{0}\left<T_\tau
  \hI(\tau)\hI(0)\right>\, e^{i\Omega \tau}d\tau \, .
\]
Here $T_\tau$ means ordering in ``imaginary time'', $\tau$,
$\beta \equiv 1/T$, temperature $T$ is measured in units of energy.
The integration over the imaginary time actually means the average
over a grand canonical ensemble with temperature $T$ and chemical
potential $\mu$.

The operators $\hat{I}^\dagger$ and $\hat{I}$ are time-dependent
interaction picture operators. Changing to Schr\"{o}dinger type
operators we write
\[
\chi_M(\Omega)= \int^\beta_{0}\left<T_\tau I(\tau)I(0)e^{\int H_T
    d\tau}\right>e^{i\Omega \tau}d\tau
\]
where $H_T$ is the tunneling Hamiltonian given by the expression
\begin{equation} \label{th-gen}
H_T(\tau)= \int \! d^2r\, d^2 r^\prime T(\br,
\br^\prime)\, [a^\dagger(\br,\tau)b(\br^\prime,\tau) + \text{h.c.}]
\end{equation}
where the integration is performed along the interface. Here $T(\br,
\br^\prime)$ is the tunneling amplitude which in general is dependent
on the coordinates both for entry to and exit from the barrier.

\begin{figure}[hb]
\begin{center}
\includegraphics[width=1.5in]{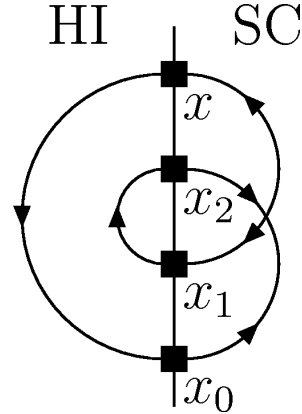}
\end{center}
\caption{Diagram relevant for the Matsubara susceptibility. \label{fig:01}}
\end{figure}

Let us first assume that
\begin{equation} \label{assump01}
T(\br,\br^\prime)=T_0\delta(\br -\br^\prime)\, ,
\end{equation}
as it was done in Ref.~\onlinecite{Kozub1}. 
Then
\begin{equation} \label{th-local}
H_T(\tau)=T_0\int d^2r[a^\dagger(\br,\tau)b(\br,\tau) +
\text{h.c.}]\, .
\end{equation} Because single electron transitions are forbidden by
the energy gap, we have to expand the expression for $\chi$ to the
second order in $H_T$. Of the many possible contractions, we are only
interested in two-electron transitions in both directions.  We get a
total of 12 different contractions. Of these, half will be only the
hermitian conjugate of the other half, and a symmetry consideration
reduces the number of relevant contractions to 3. They are
\begin{eqnarray*} 
&& \underwick{1231}{ [<1a^\dagger_\tau <2b_\tau -
a_\tau b^\dagger_\tau] [<3a^\dagger_0 >2b_0 - a_0 b^\dagger_0]
[a^\dagger_1 b_1 + >1a_1 <4b^\dagger_1] [a^\dagger_2 b_2 + >3a_2
>4b^\dagger_2] , \ (A) } 
\\&& \underwick{2132}{
[<1a^\dagger_\tau <2b_\tau - a_\tau b^\dagger_\tau] [a^\dagger_0 b_0 -
>1a_0 <3b^\dagger_0] [<4a^\dagger_1 >2b_1 + a_1 b^\dagger_1]
[a^\dagger_2 b_2 + >4a_2 >3b^\dagger_2] , \ (B1) } 
\\&&
\underwick{4132}{ [<1a^\dagger_\tau <2b_\tau - a_\tau b^\dagger_\tau]
[a^\dagger_0 b_0 - <3a_0 <4b^\dagger_0] [>3a^\dagger_1 >2b_1 + a_1
b^\dagger_1] [a^\dagger_2 b_2 + >1a_2 >4b^\dagger_2] . \ (B2) }
\end{eqnarray*}
One can show that the first one ($A$) is small at $eV \ll \Delta$,
i.e., when the single-electron transport is suppressed, while the two
others, $B1$ and $B2$, give equal contributions. We will therefore
follow only $B1$ through the further analysis.  To perform
calculations we introduce the Green's functions in a usual
way\cite{AGD}
\begin{eqnarray*}
\left<T_\tau
  b(\br,\tau)b(\br^\prime,\tau^\prime)\right>&=&F(x-x^\prime)\, , \\
\left<T_\tau
  b^\dagger(\br,\tau)b^\dagger(\br^\prime,\tau^\prime)\right>&=&
F^\dagger(x-x^\prime)\, ,  \\
\left<T_\tau
  a(\br,\tau)a^\dagger(\br^\prime,\tau^\prime)\right>&=&G(x,x^\prime)
\, .
\end{eqnarray*} 
Here $x_i\equiv (\br_i,\tau_i )$, $F(x-x^\prime)$ is
the anomalous Green function in the superconductor, and
$G(x,x^\prime)$ is the Green function of the insulator. We get
\begin{eqnarray} 
&&\left<T_\tau \hat{I}(\tau)\hat{I}(0)\right>=e^2
T_0^4\int d^2\! r\, d^2 \! r_0\, dx_1\, dx_2   \nonumber \\ 
&&\qquad \times
F(x-x_1)F^\dagger(x_2-x_0)G(x_0,x)G(x_2,x_1)\, .
\label{sp-loc}
\end{eqnarray}
Here $dx \equiv d^2\! r\, d\tau $.  We then take the
discrete Fourier transforms of the Green's functions to write them as
functions of $\Omega,\omega_i$ rather than $\tau,\tau_i$.  In this
case we will have to require that $\Omega$ and $\omega$ take discrete
values of $\Omega_m=2 \pi m T$ and $\omega_n=(2n+1)\pi T$.  Performing
the integrations over $\tau, \tau_1$ and $\tau_2$, we get
$\delta$-functions for the relations between the different discrete
frequencies, giving
\begin{eqnarray}
&&\chi(\Omega_m)=2 T e^2  T_0^4\sum_{\omega_n} \int\prod d^2r_i
\nonumber \\ && \quad \times
F(\br-\br_1,\omega_n) F^\dagger(\br_0-\br_2,-\omega_n)
\nonumber \\ && \quad \quad \times
G(\br_0,\br,\omega_n-\Omega_m)G(\br_2,\br_1,-\omega_n)\, .
\end{eqnarray}
The Feynman diagram corresponding to this expression is shown in
Fig.~\ref{fig:01}.

We assume the localized states to have hydrogen-like wave functions,
centered on impurities in positions $\br_s$, with energies
$\epsilon_s$ and localization length $a$,
\begin{equation} \label{wf1}
\Psi_s(\br)=(\pi a^3)^{-1/2}\, e^{-|\br-\br_s|/a}
\end{equation}
and the insulator Green's function is
\begin{equation} \label{wf2}
G(\br,\br^\prime,\omega_n) = \sum_s
\frac{\Psi_s^*(\br)\Psi_s(\br^\prime)}{i\omega_n-\epsilon_s}\, .
\end{equation}
For the anomalous Green's function we use
\begin{eqnarray} \label{wf3}
F(R,\omega_n)&=&\int \frac{d^3p}{(2 \pi
  \hbar)^3}\frac{\Delta}{\Delta^2+\xi_p^2+\omega_n^2}
e^{-i\mathbf{p}\cdot\mathbf{R}/\hbar}  \nonumber \\
&=&\frac{\pi g_m \Delta}{2 \sqrt{\Delta^2 +
    \omega_n^2}}\frac{\sin(Rk_F)}{Rk_F}\, e^{-\frac{R
    \sqrt{\Delta^2+\omega_n^2}}{ \pi \xi  \Delta }}\, .
\end{eqnarray}
Here $\xi_{\bf{p}}=(p^2-p_F^2)/2m$, $\xi$ is the coherence length in a
superconductor, $g_m=mp_F/\pi^2$ is the density of states in a metal.

\section{Calculations} \label{calculations}

So far we just followed Ref.~\onlinecite{Kozub1}, but we will now
demonstrate how the oscillations of the anomalous Green's functions,
$\propto \sin Rk_F$, lead to a significant decrease of the result
comparing to the simple estimates presented there. It turns out that these
oscillations give rise to pronounced cancellations in the
susceptibility for the case of a hydrogen-like impurity state.

We now define $\br_s$ and $\br_l$ as coordinates of the impurities
contributing to $G(\br_2,\br_1)$ and $G(\br_0,\br)$,
respectively. Then the Matsubara susceptibility can be expressed as
\begin{widetext}
\begin{eqnarray} \label{chiM}
 \chi_M (\Omega_m)&=& \frac{T e^2
  |T_0|^4 g_m^2}{2a^6} \sum_{sln} \frac{
  \Delta^2}{\Delta^2+\omega_n^2}
\frac{I^2_{sl}}{(-i \omega_n-\epsilon_s)(i \omega_n -i\Omega_m -
  \epsilon_l)}\, , \\
I_{sl}&=&\int d^2\!r\, d^2\!r_1 \, \frac{\sin k_F
  |\mathbf{r}-\mathbf{r}_1|}{k_F|\mathbf{r}-\mathbf{r}_1|}\,
\exp \left(-\frac{|\br_1-\br_s|+|\br -\br_l|}{a}
-\frac{(|\br-\br_1|)\sqrt{\Delta^2+\omega_n^2}}{\pi   \xi \Delta} \right)
 \, .  \label{I01}
\end{eqnarray}
\end{widetext}

\begin{figure}[ht]
\centerline{
\includegraphics[width=5cm]{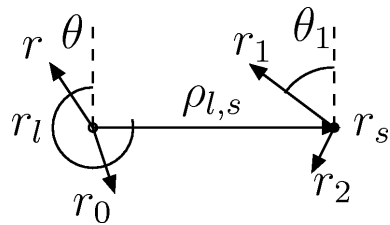}
}
\caption{Coordinates for the spatial integration. \label{coord}}
\end{figure}

It is safe to assume that $|\br-\br_1|$  and $|\br_0-\br_2|$ are of
the order the distance $\rho_{sl}$ between the impurities projected
onto the interface, but with variations of the order $a$, where $k_F a
\gg 1$. The localization length $a$ can be estimated as the Bohr
radius  $a_0=4 \pi \kappa  \hbar^2/m^* e^2$. Assuming $m_* \approx 0.1
m_e$ where $m_e$ is the mass of a free electron, and $\kappa \approx
10$, we get $k_F a \approx 100$. Since the superconductor localization
length is much greater, $\xi \gg a$, we can safely replace
$|\br-\br_1| \to \rho_{sl}$ in Eq.~(\ref{I01}) in all places
\textit{except} the $\sin k_F |\br-\br_1|$, which oscillates rapidly.
Choosing the coordinates as shown in Fig.~\ref{coord} we obtain
\begin{eqnarray*} I_{sl} &=& \frac{1}{k_F
\rho_{sl}}e^{-\frac{\rho_{sl}}{\pi
\xi}\frac{\sqrt{\Delta^2+\omega_n^2}}{\Delta}} \\ && \times \int d^2\!
r d^2\! r_1 \, \sin k_F |\br-\br_1| \, e^{-(|\br_1-\br_s|
+|\br-\br_l|) /a}\, .
\end{eqnarray*}

As $\br$ and $\br_1$ are located in the interface plane, it is natural
to choose the origins for $\br$ and $\br_1$ to be at the projections
of $\br_s$ and $\br_l$, respectively, into the interface plane of
the. Then $ d^2\! r_i = r_i\, dr_i\, d\theta_i$,
\begin{eqnarray*}
&&
|\br-\br_l|=\sqrt{r_1^2+z_l^2}, \ |\br_1-\br_s|=\sqrt{r_1^2+z_s^2}\, ,
\ |\br_1-\br_2|
\\&&
=\sqrt{(\rho_{sl}+r \sin \theta - r_1 \sin \theta_1)^2 + (r
    \cos \theta - r_1 \cos \theta_1)^2}
\end{eqnarray*}
where $z_s$ and $z_l$ are the distances from the impurities to the interface.

The minimal distance between the impurities taking part in the
coherent transport, $\rho_{sl}^{\min}$,  is limited by Coulomb
correlation. If $a/\rho_{s,l}^{\min} << (k_F a)^{-1}$ one can neglect
the items containing $\cos \theta_i$ in the above expression. Then
$|\br-\br_1| \approx \rho_{sl}+r \sin \theta - r_1 \sin \theta_1$, and
the integrals over $\br$ and $\br_1$ can be calculated
exactly using the formula
\begin{eqnarray*}
&&\int_{-\pi}^\pi d\theta \sin (k_F r \sin \theta + \phi)
=\Im \int_{-\pi}^\pi \! \! d\theta \, e^{i k_F r \sin \theta + i \phi} \\&&
=\Im e^{i\phi}\int_{-\pi}^\pi d\theta e^{i k_F r \sin \theta}
=  2 \pi \sin(\phi) J_0(kr)\, .
\end{eqnarray*}
After integrating over both angles we get
\begin{eqnarray} \label{ir1}
I_{sl}&=& \frac{(2 \pi)^2 \sin k_F \rho_{sl}}{k_F
  \rho_{sl}}e^{-\frac{\rho_{sl}}{\pi
    \xi}\frac{\sqrt{\Delta^2+\omega_n^2}}{\Delta}} I_r(z_s)I_r(z_l)\,
  , \nonumber \\
I_r(z)&=&\int_0^\infty rdr  J_0(k_F r) e^{-\sqrt{r^2 + z^2}/a}\, .
\end{eqnarray}
The integral $I_r$ will be referred to extensively in later sections,
as most of the modifications we will discuss change this integral only,
leaving all other calculations unaltered.
We can then use the following identity~\cite{Prudnikov}
\begin{eqnarray*}
&&\int_0^\infty x dx\, e^{-p\sqrt{x^2 + z^2}}J_0(c x) \\
&& \quad =
p(p^2+c^2)^{-3/2}\left(1+z\sqrt{p^2 + c^2}\right)e^{-z\sqrt{p^2 + c^2}} \\
&&\quad \quad [\Re (p) > |\Im (c)|; \quad \Re (z) > 0] \, .
\end{eqnarray*}
Combining all integrals,
this gives:
\begin{eqnarray*}
I_{sl}&=&\frac{(2 \pi)^2 a^4\sin k_F \rho_{sl}}{k_F
  \rho_{sl}}\, \frac{\cF(z_l)\cF(z_s)}{(1+k_F^2
  a^2)^{3}}\, e^{-\frac{\rho_{sl}}{\pi
    \xi}\frac{\sqrt{\Delta^2+\omega_n^2}}{\Delta}}\, , \\
\cF(z)&=&\left(1+\frac{z}{a}\sqrt{1 + k_F^2 a^2}\right)\, e^{-\frac{z}{a}\sqrt{1 + k_F^2 a^2}}
\end{eqnarray*}
Using the assumption that $k_F a \gg 1$ the function $\cF(z)$
 simplifies to
$$\cF(z) \approx(1+k_Fz)e^{-zk_F}\, , $$
The essential observation here is that the expression
for $I_{sl}$
contains a factor $e^{-(z_l+z_s) k_F}$, which again should be squared for
the final result.
This conclusion contradicts
Ref.~\onlinecite{Kozub1}, where the factor  $e^{-(z_l+z_s)/a}$ was
predicted. The strong decay of $I_{sl}$ as a function of $z_l$ and
$z_s$ means that only  pairs very close to the interface
can contribute to the Cooper pair transfer. 
This means that the theory as presented above and in Ref.~\onlinecite{Kozub1} 
proves its own inadequacy, since the assumption that the  
wave functions of the localized states is 
bulk hydrogen-like ones requires $z_l$ and $z_s$ to be at least of the order
of $a$. For closer impurities, the wave function is modified by the 
vicinity of the surface, and the result becomes strongly dependent on
unknown properties of the surface states. In this paper we will not 
consider these close impurities, but discuss how details of the 
barrier may change the above result back to the $e^{-(z_l+z_s)/a}$
of Ref.~\onlinecite{Kozub1} and thus allow the main contribution 
to come from pairs further from the barrier. 

Note also the extreme accuracy to which the positive and negative 
contributions to $I_r(z)$ cancel. Between two adjacent zeroes of 
$J_0(k_Fr)$ the integral is of order 1 for small $r$ (for $r>a$ it 
gets damped by the exponential), yet the final integral is of 
order $e^{-k_Fz}\approx e^{-k_Fa} \approx 10^{-44}$ if $k_Fa=100$.

A closer analysis of the integral over $r$ shows that it
accumulates a negative value of the order $e^{-z/a}$ for
small $r$, which is almost canceled by an equivalent positive contribution
for very large $r$. The cancellation is found to be strongly dependent on
the exact shape of the wave function. Consequently, one may conclude
that the cancellation can be lifted by specific properties of the
tunneling amplitude, which could introduce an effective cut-off of the
integration over $r$.
In the following Sec.~\ref{linearconductance} several models will be
discussed, where the importance of hitherto ignored details in the
tunneling barrier will be made clear, and the importance of the
assumed wave function for the localized states will be discussed.

To complete the calculation of the conductance, we must now perform
the summation over the Matsubara frequencies in the standard way,
replacing
\[
T \sum_{\omega_n} f(i \omega_n) = \oint \frac{d \epsilon}{4 \pi i}
f(\epsilon) \tanh \frac{\epsilon}{2 T}.
\]
Under the assumptions we have made, this integration is independent 
of the details of the
spatial integration, and will not be affected by the modifications
introduced in later calculations. Integrating over the contour shown
in Fig.~\ref{contour}, we get an expression for conductance depending
on the discrete variable $\Omega_m$. Making an analytical continuation
of this function, and taking the limit as $\Omega_m$ goes to zero, we
finally get an expression for the conductance:
\begin{figure}[h]
\centerline{
\includegraphics[width=5cm]{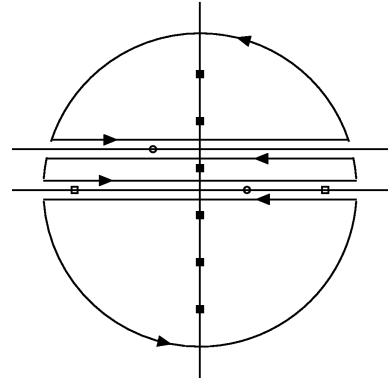}
}

\caption{Integration contour. The values $\epsilon=\epsilon_s$ and
  $\epsilon_l-i\Omega_m$ are shown by circles ({\large $\circ$}),
 the
 values $\epsilon=\pm \Delta$ are shown by empty squares ($\square$),
 while the poles of $\tanh(\epsilon/2T)$ are shown by filled squares.
 \label{contour}}
\end{figure}
\begin{eqnarray}
G&=& \frac{\pi(2 \pi)^4a^2  e^2 |T_0|^4 g_m^2}{2(k_F a)^{12}
  T}\sum_{s\ne l} \frac{\sin^2 (k_F \rho_{sl})}{(k_F \rho_{sl})^2}\,
  e^{-2\frac{\rho_{sl}}{\pi \xi}}
\nonumber \\ && \times
  \cF(z_s)\cF(z_l)n(\epsilon_l)n(\epsilon_s)\delta(\epsilon_s +
  \epsilon_l)\, .
\end{eqnarray}
Here $n(\epsilon)=(e^{\epsilon/T}+1)^{-1}$ is the Fermi distribution.

As shown in Ref.~\onlinecite{Kozub1}, it is important to include the
effect of Coulomb interaction between the occupied sites which results
in additional energy $U_C=e^2a^2\kappa \rho_{sl}^3$. However, 
accounting for this interaction does not change the strong
cancellation. We will come back to the role of Coulomb interaction in
Sec.~\ref{nonlinearconductance} where we discuss non-Ohmic conductance.  In
the next section we discuss how robust the cancellation is  and how it
is influenced by the properties of the tunneling barrier.

\section{Ohmic Conductance} \label{linearconductance}

To understand how robust the cancellation shown in the previous
section is, we will analyze several aspects of Ohmic transport through
the interface.

\subsection{Importance of Impurity Wave Function}

The hydrogen-like wave function is a typical approximation for shallow
centers in semiconductors. However, the crossed Andreev transport can
also take place in mesoscopic devices, e.g., between a bulk
superconductor and a pair of quantum dots. This is, in particular, the
case for the previously suggested spin entangler.\cite{spin-entangler}

If the above cancellation is correct also for that case, then the
crossed Andreev transport would be hardly feasible since the dots would
have to be located virtually at the interface. However, the tails of
the wave functions of the electrons localized at quantum dots are far
from being hydrogen-like. In general, they are dependent on the design
of the quantum dots. In particular, for the lateral quantum dots
designed by a properly engineered gate potential one can expect
parabolic confinement. In this case the wave function tail is Gaussian
rather than exponential.

To check whether the above cancellation exist in this case we have
repeated the calculations of Sec.~\ref{calculations},  but replacing
 $\Psi_s(\br)$ with
$\Psi_{G}(\br)=(2\pi a^2)^{-1/2}e^{-|{\bf r-r}_s|^2/2a^2}$.
As a result, the contribution of a given pair decays as
$e^{-(z_s+z_l)/a}$, and one returns to the estimates of
Ref.~\onlinecite{Kozub1}. Thus the design of quantum dots chosen for
the spin entangler can be crucial for it's potential success.

\subsection{Importance of Barrier Properties}

In Sec.~\ref{calculations} we assumed that the tunneling amplitude is
  \textit{local} and \textit{coordinate-independent},
  $T(\br,\br')=T_0\delta(\br-\br')$. It means that during tunneling an
  electron can transfer its momentum to some disorder-induced
  scatterers, and the tunneling amplitude is independent of the
  incident angle.
To study the role of this simplification, we will proceed as follows. First, we consider the case of an
\textit{ideal} barrier for which the tunneling amplitude depends only
on the incident angle. We will show that such dependence does not
remove the cancellation, and the decay $\propto e^{-(z_l+z_s)k_F}$
persists. Then we will consider the case of a barrier with
fluctuating strength. We will find that fluctuations of a scale
$\lesssim a$ can strongly facilitate
transport restoring the $e^{-(z_l+z_s)/a}$ dependence.

\begin{figure}[h]
\begin{center}
\begin{minipage}[c]{4.1cm}
\centerline{\includegraphics[width=4cm]{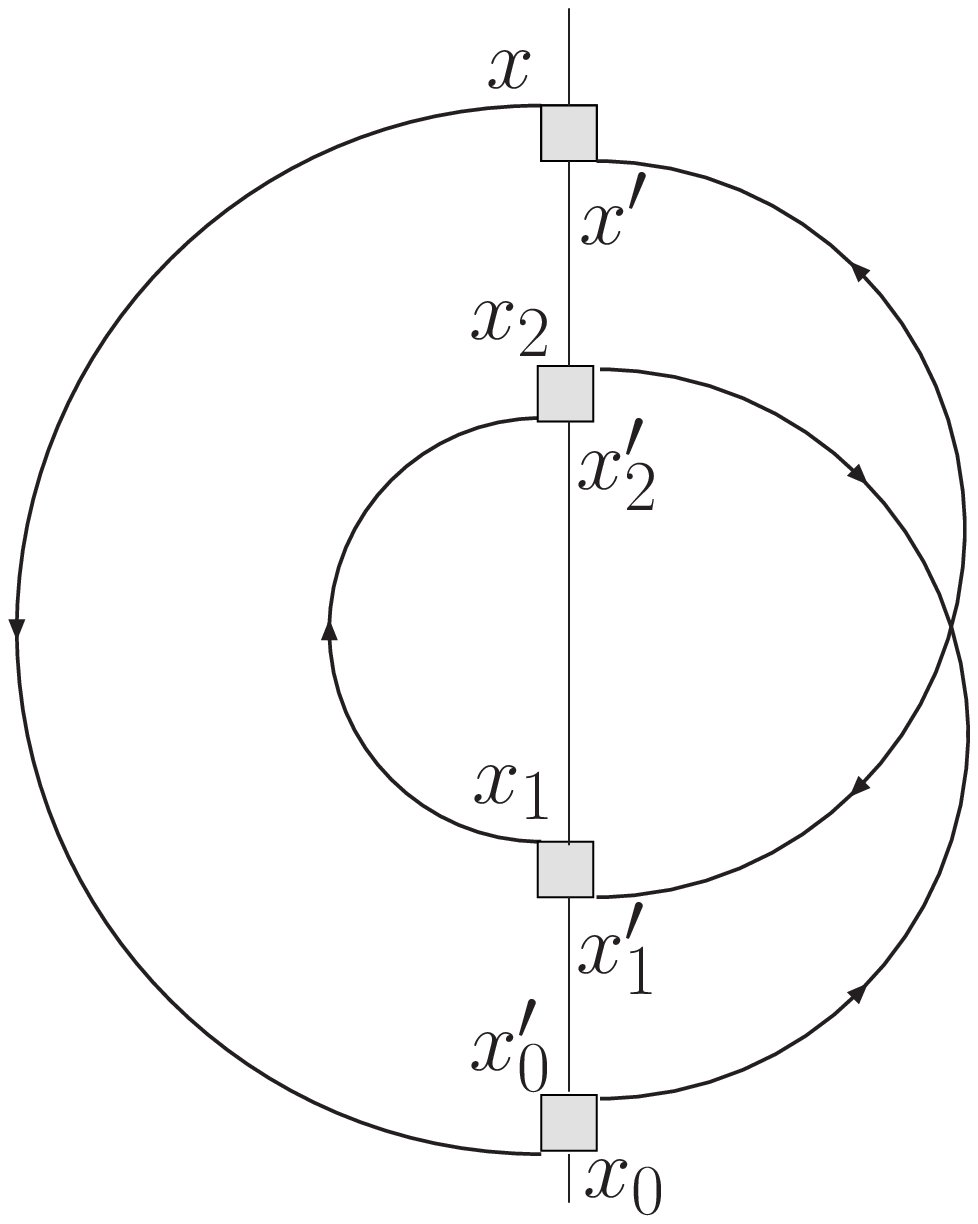}}
\end{minipage} \hfill
\begin{minipage}[c]{3.9cm}
\centerline{\includegraphics[width=3.8cm]{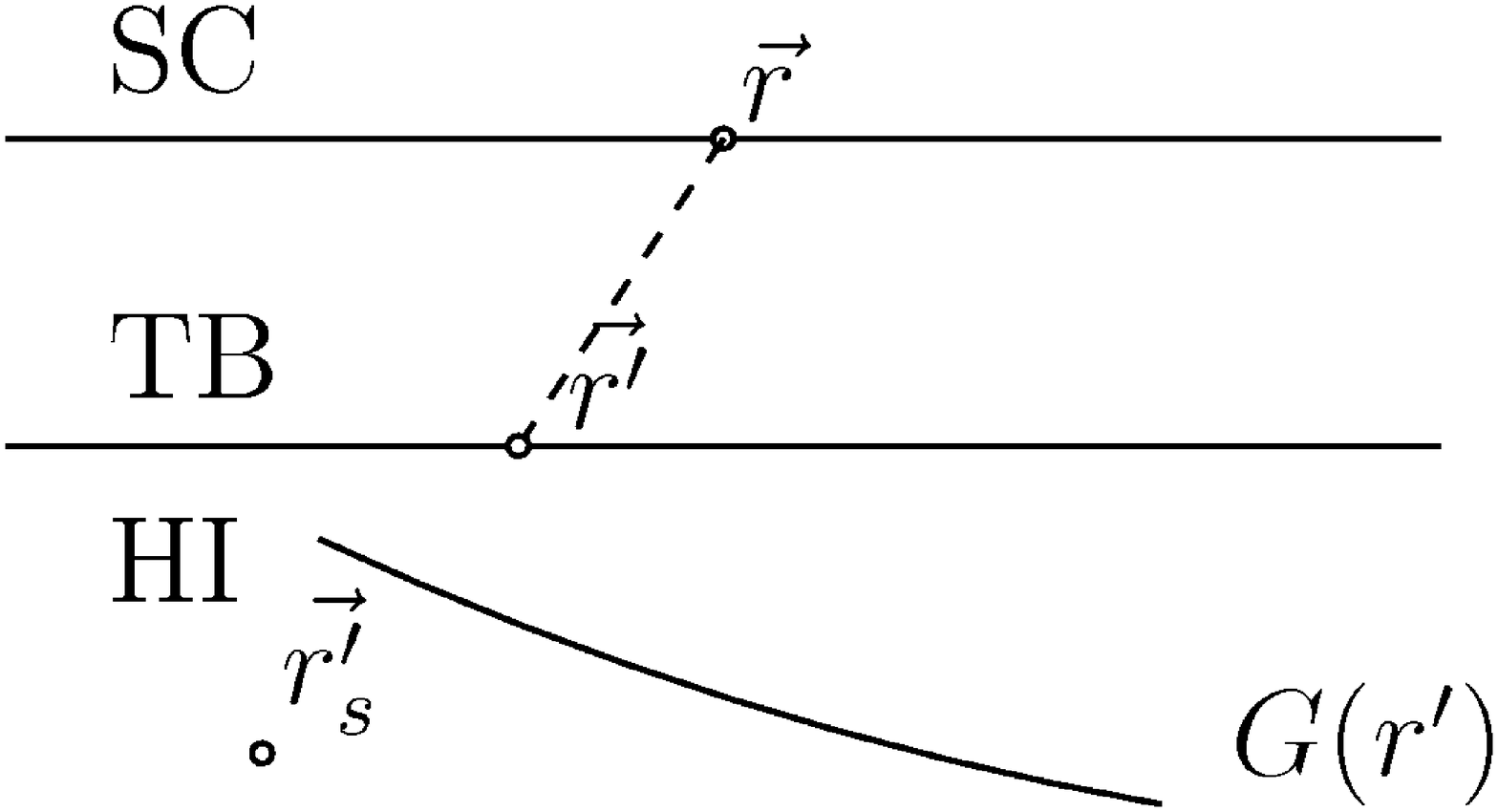}}
\end{minipage}
\end{center}
\caption{Left panel: The diagram of Fig.~\protect{\ref{fig:01}} for
the case of nonlocal tunneling transparency. Right panel: More
detailed sketch of the coordinates. \label{fig:di-nl}}
\end{figure}

To make these consideration more specific let us assume that the
effective barrier thickness, $d$, fluctuates along the interface. Then
the tunneling amplitude is nonlocal, and the tunneling Hamiltonian
acquires the general form of Eq.~(\ref{th-gen}). Consequently, the
coordinates of the Green's functions for HI- and SC-side are
different, and the proper diagram has the form of Fig.~\ref{fig:di-nl}
rather than that of Fig.~\ref{fig:01}.
 For the following it is convenient to normalize the
tunneling amplitude to $T_0$,
$$T(\br,\br^\prime) \equiv T_0\, f(\br,\br^\prime)\, .$$
 Then the spatial integral can be written as [cf. with Eq.~(\ref{sp-loc})]

\begin{eqnarray} \label{sp-nonloc} &&\int \! \prod  d^2\! r_i \, d^2
\!  r_i^\prime \, f(\br,\br^\prime) f(\br_0,\br^\prime_0)
f(\br_1,\br^\prime_1)f(\br_2,\br^\prime_2) \nonumber \\ &&  \quad
\times
\sum_{\omega_n}F(\mathbf{r}-\mathbf{r}_1,\omega_n)F^\dagger(\mathbf{r}_0
- \mathbf{r}_2,\omega_n) \nonumber \\ &&\quad \quad  \times
G(\mathbf{r}^\prime,\mathbf{r}_0^\prime,\omega_n-\Omega_m)
G(\mathbf{r}_2^\prime,\mathbf{r}_1^\prime,\omega_n)  \, .
 \end{eqnarray}
Following the previous calculations, this can be split into separate,
identical integrations for each impurity.  To separate the roles of
barrier thickness fluctuations and dependence of the incident angle
let us express the tunneling amplitude as
\begin{equation}
f(\br,\br^\prime)=g(\br)h(\br^\prime-\br).
\end{equation}
where $g(\br)$ describes spatial variations in the barrier, while
$h(\br^\prime-\br)$ accounts for the dependence on the incident
angle. Here both vectors $\br$ and $\br^\prime$ belong to the
interface plane. The function $h(\br^\prime-\br )$ can be assumed to
depend only on  $| \br^\prime-\br |$.

\paragraph{Smooth Barrier:}

Let us start with the case when $g(\br)=g_0= \text{constant}$. The
basic spatial integration is
\begin{equation}
I(r,r_0,\omega_n) \equiv \int d^2r^{\prime \prime} G(|\br + \mathbf{r}^{\prime
  \prime}|,r_0^\prime,\omega_n)h(r^{\prime \prime})\, .
\end{equation}
Here we have taken into account that the Green's
function $G(\br,\br_1, \omega_n)$ depends only on $\sqrt{z^2+r^2}$ and
$\sqrt{z^2+r_1^2}$. Let us now assume that $h(r^{\prime \prime})$
decays much more rapidly than $G$. That allows us to expand the
integrand in powers of $x^{\prime \prime}$ and $y^{\prime \prime}$
keeping only the second order (the first order term vanishes on integration),
$$G(|\br + \mathbf{r}^{\prime  \prime}|,r_0^\prime,\omega_n) \approx
 G  +\frac{x^{\prime \prime 2}}{2}\frac{\partial^2 G}{\partial x^2} +
\frac{y^{\prime \prime 2}}{2}\frac{\partial^2 G}{\partial y^2}$$ with
$G \equiv G(r,r_0^\prime,\omega_n)$. Now let us consider the simplest
case of a rectangular barrier for which the function $h(r)$ can be
modeled as $h(r)=d^{-2}e^{-B(r/d)^2}$ where $B \gtrsim 1$ is some
dimensionless constant. This model follows from an assumption that the
barrier is uniform along the surface and rectangular. Then the
tunneling exponent can be written as
\begin{eqnarray*}
&& -d \sqrt{2m[U-E(1-k_\parallel^2/k^2)]}
\approx - \lambda_0
-B (r/d)^2\, ,
\end{eqnarray*}
where $ \lambda_0 \equiv d\sqrt{2m(U-E)}$, $B\equiv \lambda_0
E/2(U-E)$, $U$ is the barrier height, while $E$ is the electron
energy.
Using this model and explicit
expression~(\ref{wf2}) for the Green's function we obtain
\[
\begin{split}
&I(r,r_0,\omega_n) =  G(r,r_0^\prime,\omega_n) \,c(r)\, ,
\\
 c(r) \equiv \frac{\pi g_0}{B}&\!\!\left[1+ 
   \frac{d^2}{4Ba^2}\!\! \left(\!\!\frac{r^2}{r^2+z^2}-\frac{2
       a}{\sqrt{r^2 + z^2}}+\frac{a r^2}{\sqrt{r^2 + z^2}^3}\!\!
   \right) \! \right] \! . \label{tmp1} \nonumber
\end{split}
\]
The integral $I_r$, Eq. (\ref{ir1}), now changes to
\[
I_r(z)=\int_0^\infty rdr  J_0(k_F r) e^{-\sqrt{r^2 + z^2}/a} c(r)\, .
\]
At $d \ll a$ the correction induced by the angular dependences of the
transmission is small, but, in principle, it could be sufficient to
lift the cancellations that give the $e^{-k_F z}$-dependence. However,
a combination of analytical and numerical analysis shows that the
additional terms also lead to the  $e^{-k_F z}$-dependence.  We
therefore conclude that the dependence of the tunneling transparency
of a uniform barrier on the incident angle still leads to the
$e^{-k_F z}$-decay of the crossed Andreev transport.

\paragraph{Inhomogeneous barrier:}

We now turn to the situation were we have fluctuations in the barrier
strength that facilitate tunneling through the places where the
barrier is thin. Let us assume that the typical size of these regions,
$\lambda$ is much less than the localization length, $a$, but larger than
$k_F^{-1}$,
$$k_F^{-1} \ll \lambda \ll a\, .$$
Now
we cannot assume $g(\br)$ to be constant over the region spread by the
impurity potential. For
  simplicity, we will in this case write
  $h(\br^{\prime\prime})=\delta(\br^{\prime\prime})$, as we have
  previously shown that these corrections do not change the principal
  behaviour of the transport.
The shape of $g(\br)$ is dependent on the roughness, and on the
relative positions of the barrier minimum and
the impurity center. For a barrier with a parabolic
minimum one can show that $g(r)$ has a Gaussian shape,
\[
g(r)=e^{-r^2/a^2\sigma} \, .
\]
The general analysis of this situation is complicated, as the impurity
center may not coincide with the center of the barrier minimum. Two
simplified cases are still sufficient to shed light on the
situation.

Let us for simplicity start with the case when the minimum in the
barrier strength coincides with the projection of the impurity center
on the interface. In this case the integral for $I_r (z)$ similar to
Eq.~(\ref{ir1}) can be analyzed in detail. It turns out that with
increasing $z$ it crosses over from   $e^{-k_F z}$ to $e^{-z/a}$ at
some $z^\ast$ which depends on $\sigma $. The quantity $z^\ast$
decreases with decrease of $\sigma$, $z^\ast=a$ at $1/\sigma=0.15$.  Thus
the barrier inhomogeneity (modeled by small $\sigma$) facilitates the
transport by eliminating the cancellation.
\begin{figure}[h] \centering
\includegraphics[width = 7cm]{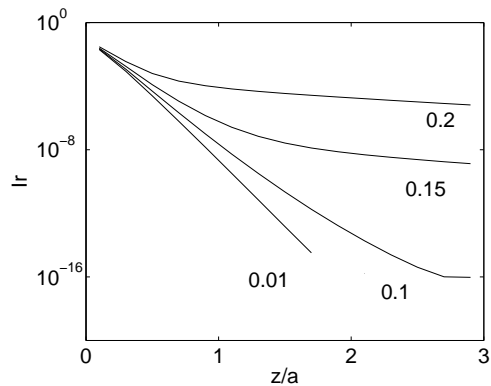}
\caption{The dependence $I_r(z)$ for $k_Fa=20$ and  $1/\sigma =0.01,
0.1, 0.15, 0.2$.  }
\label{gammaz}
\end{figure} Now we can relax the previous assumption that the barrier
strength minimum occurs exactly at the projection point of an impurity
center and consider the situation where the minimum is off-center, but
varies so fast that $G(r)$ can be considered constant in
comparison. In this case we can replace $G(r)=G(r_{\min})$, $r_{\min}$
being the barrier minimum coordinate, and move the origin of the
integration to $r_{\min}$. The integral will then again give a simple
$e^{-z/a}$-dependence. Thus we see that the cut-off introduced by a
clear minimum in the barrier is sufficient to change the
$z$-dependence of the conductance per impurity.

If there are several minima within one single impurity, one could
imagine that these could give new interference effects. However,
integration over several minima corresponds to simply summing up the
contribution from the separate integrals. Each minimum will be coupled
to a minimum on another impurity, giving a prefactor of $\sin^2 \!
\rho_{sl}$. When several sines are added at each impurity, we get
$$ (\sin \rho_{sl}+\sin \rho_{s'l'})^2 
=\sin^2 \! \rho_{sl} +\sin^2 \! \rho_{s'l'} +2\sin \rho_{sl} \sin
\rho_{s'l'}.$$ When averaging over several pairs, the $\sin^2$-terms
will survive, while the cross-terms will average to zero. We therefore
assume several minima within the range of each impurity to be
equivalent to several separate pairs in the total
averaging. Considering the probability of a pair accepting a
Cooper-pair, the effect on one minimum of the pair already being
occupied due to another minimum should be negligible.

The characteristic localization length of the electron under the
barrier, we call $\alpha$. In order for these considerations to be
relevant, the barrier thickness $d$ must vary with several $\alpha$
over a length scale much shorter than $a$. Changes in barrier
thickness that do not meet this condition are better analyzed in terms
of the following model.

\paragraph{Barrier with a block-like disorder:}

\begin{figure}[h]
\centering
\includegraphics[width = 7cm]{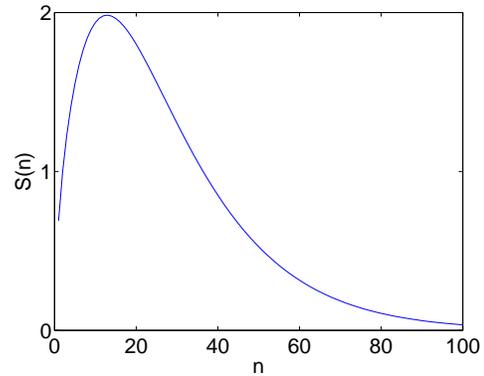}
\caption{$S(n)$, the integral cut-off after $2n$ zeroes of the Bessel
function, $k_Fa=100, z/a=1$  \label{accpairs}}
\end{figure}

The model discussed above relies on a change in barrier thickness that 
is of the order of $\alpha$. Because of the very fine cancellations 
in the integral $I_r$, Eq. (\ref{ir1}), much smaller changes in barrier 
height can be important, provided they are on a length scale of the 
order of $a$. This can be demonstrated using 
a simple model based of the analysis of the integral 
\[
I_r(R,z) \equiv \int_0^R rdr \, J_0(k_F r) e^{-\sqrt{r^2 + {z}^2}/a}
\]
as a function of the cut-off $R$. Obviously, $I_r(z)=\lim_{R\to
  \infty} I_r(R,z)$. By splitting the integration over $x$ into
intervals divided by the zeroes of the Bessel function, we get 
successive contributions of alternating signs and close absolute
values. The result
approaches the $e^{-k_F z}$ behaviour seen before when $X$ goes to
infinity. The absolute value of the sum, $S(n,z)$  of an even number of the
intervals defined as 
\begin{eqnarray*}
S(n,z)&=&\sum_{m=0}^{2n} \int_{R_m}^{R_{m+2}} rdr \, J_0(k_F r)
e^{-\sqrt{r^2 + {z}^2}/a} \\
&=&\int_{0}^{R_{2n+2}} rdr \, J_0(k_F r) e^{-\sqrt{r^2 + {z}^2}/a}
\end{eqnarray*}
where $R_m/a$ is the $m$'th zero of the Bessel function, $R_0=0$, will
therefore have a maximum for some $n$, as shown in Fig.~\ref{accpairs}.
 This maximal value of $S(n,z)$ this maximum varies with
$z$ as $e^{-z/ a }$, and  the corresponding cut-off radius, $R^\ast
\equiv R_{n_{\max}}$,  is a slowly  varying function of the ration
$z/a$, corresponding to 20 to 100 zeroes of the Bessel function.

Based on this property we construct a simple model of a barrier with a
block-like disorder assuming
\begin{equation} \label{mg} g(r)=1+\eta \, \Theta(R^\ast -r)
\end{equation} where $\Theta (r)$ is the Heaviside
step-function. Since this contribution only slowly decays with $z$,
even a small barrier variation, $\eta$, can give significant
contributions. Since the first term in Eq.~(\ref{mg}) leads to a decay
$\propto e^{-k_Fz}$ while the second contribution decays $\propto
e^{-z/a}$ we only need $\eta >e^{-(k_F-1/a)z}$.
\begin{figure}[h] \centering
\includegraphics[width = 7cm]{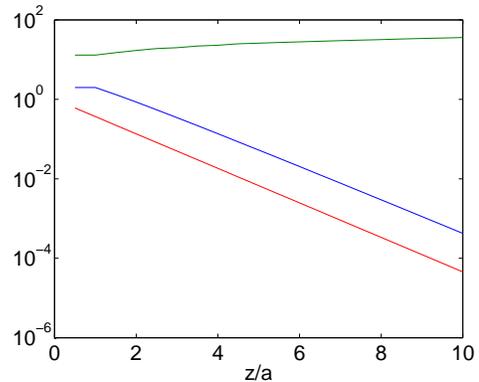}
\caption{The functions $n_{max}(z)$ (top line) and $I_r(R^\ast,z)$
(middle line) for $k_Fa=100$. The bottom line shows the function
$e^{-z/a}$.
\label{cutoffint}}
\end{figure} Figure \ref{cutoffint} shows the optimal cut-off radius,
$R^\ast$, as a function of  $z$, as well as $I_r(R^\ast,z)$ compared
to a graph showing $e^{-z/a}$.

Obviously, small variations in inhomogeneity size, $R$, will give
large variations in the result, this can be remedied by smearing the
cut-off over a period or 
two of the Bessel function. If we assume the center of the barrier
reduction to be slightly displaced from the impurity center, the
integration over angles can  probably be found to be some such smeared
step function. At some point, the smearing will be such that the
cancellation is no longer lifted, and the barrier minimum gives
insignificant contribution.

Thus we see that a small variation in the barrier
thickness, as long as it is at the proper length scale and 
centered on the impurity center, can drastically reduce the interface 
resistance of the barrier.

\section{Non-Ohmic Conductance}\label{nonlinearconductance}

Having established that there exists a range of applicability of the
underlying model we now address the non-Ohmic behavior of the
interface conductance. To figure out the nonlinear properties one has
to compare $eV$ with other relevant energy scales: temperature $T$,
inter-site  Coulomb repulsion energy $U_C$, and the energy splitting
$\epsilon_s-\epsilon_l$ of a pair. In addition, we have to consider
the width of the $\delta$-function in energy that selects which pairs
may contribute. This width can be estimated as a typical inverse life
time of an electron at a localized site forming the pair.  We always
assume $eV\ll\Delta$ excluding in this way the possibility of
single-electron transport.

The net current over the barrier can be seen as the difference between
current from the superconductor to the insulator, $I_{S\to I}$ and
current from insulator to superconductor $I_{I\to S}$. These currents
are in turn determined by the transition probability and the
occupation probabilities for the involved states.  Following
Ref.~\onlinecite{GGP}  we will assume all matrix elements to be
energy-independent. Thus the only variations in the transition rate is
due to the occupation probabilities.  We must also remember the
eV-dependence of the $\delta$-function, describing the conservation of
energy, which regulates which pairs contribute to the transport.

Since the superconducting condensate has a macroscopic number of
states at the level of the chemical potential, the current $I_{S\to
I}$ is only dependent on the probability of finding an empty pair in
the insulator, while $I_{I\to S}$ requires an occupied pair. In both
cases the relevant pair will have to satisfy energy conservation.

To keep track of realistic situations we assume that the entire
voltage drop occurs at the barrier, but allow for a small portion of
the insulator near the interface to be filled up or emptied by
electrons due to the voltage drop over the barrier. This region models
the depletion zone of a semiconductor heterojunction. For impurities
outside the depletion zone, the Fermi level is assumed to be fixed
relative to the impurity energy levels. In this case the
$\delta$-function in energy must be chosen as
$\delta(\epsilon_s+\epsilon_l+U_C-2eV)$, and the occupancy numbers are
given as before, as $n(\epsilon)$. Very close to the barrier, inside
the depletion layer, we instead use a picture where we keep the
impurity energy levels constant relative to the superconductor
condensate, but adjust the Fermi level to get
$n(\epsilon,eV)=n(\epsilon+eV)$.
\begin{figure}[hb] \centerline{
\includegraphics[height=2.3cm]{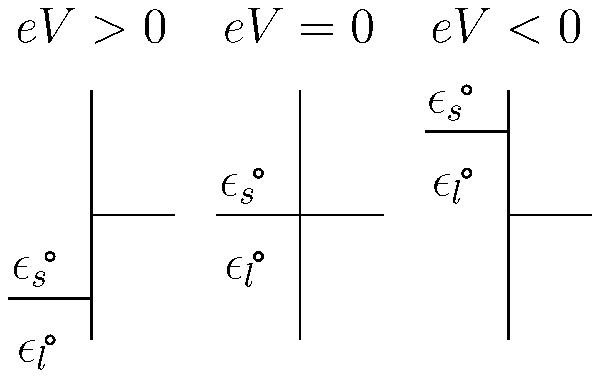} \hfill
 \includegraphics[height=2.3cm]{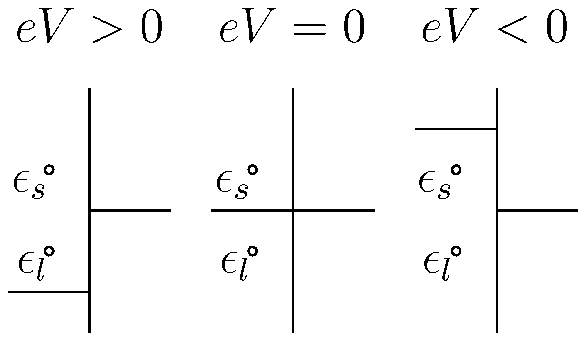} }
\caption{Electrochemical potential in HI (left) and SC (right) outside
(left panel) and inside (right panel) the depletion zone
\label{fig:ecp01}}
\end{figure}

\begin{figure}[h] \centerline{
\includegraphics[width=8cm]{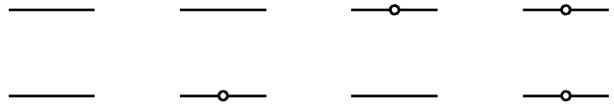} }
\caption{Energy levels for 4 configurations of the pair:
$\epsilon_1=0$, $\epsilon_{10}=\epsilon_s$,
$\epsilon_{01}=\epsilon_l$,
$\epsilon_{11}=\epsilon_s+\epsilon_l+U_C$. \label{fig:terms1}}
\end{figure}

In both cases we have to consider the Coulomb energy $U_C$, and the
simplest way of accommodating it is by describing each pair as a
four-level system corresponding to the 4 following configurations: (i)
both sites are empty, (ii, iii) one site is occupied, and (iv) both
sites are occupied. The configurations are shown in
Fig.~\ref{fig:terms1}.
 For simplicity we disregard Coulomb interaction with
charges outside the pair. We can then write a partition function for
the four-level system, and use this to find the probabilities of a
pair being empty - allowing a Cooper pair to fill it - or if filled,
allowing one Cooper pair to be created.  Using the energies defined in
the figure, the partition function can be written as
\[ Z= 1+ e^{-(\epsilon_s-\mu)/T} + e^{-(\epsilon_l-\mu)/T} +
e^{-(\epsilon_s+\epsilon_l-2\mu+U_C)/T}
\]
and the probabilities of the different configurations in similar notation:
\begin{eqnarray*}
 P_{00}=Z^{-1}, &\quad&  P_{10}=Z^{-1} e^{-(\epsilon_s-\mu)/T}, \\
P_{01}=Z^{-1}e^{-(\epsilon_l-\mu)/T}, &\quad&
P_{11}=Z^{-1}e^{-(\epsilon_s+\epsilon_l-2\mu+U_C)/T}.
\end{eqnarray*}
Here  $\mu$ is the chemical
potential in the superconductor.

The current through the interface is proportional to the difference
\begin{equation} P_{00}-P_{11}
=Z^{-1}\left(e^{-(\epsilon_s+\epsilon_l-2\mu+U_C)/T}-1 \right)\, .
\end{equation} With the inclusion of the Coulomb energy this can no
longer be factorized into separate occupation probabilities of the two
impurities of the pair, but if we set $U_C=0$, it can be seen that the
difference simplifies to the former results.

If we consider the pairs inside the depletion zone, we use $\mu=-eV$.
Taking into account the energy conservation law requiring
$\epsilon_s+\epsilon_l +U_C =0$ one can express the difference
$P_{00}-P_{11}$ as
$$
\frac{1-e^{2eV/k_BT}}{1+e^{(-\epsilon_s+eV)/k_BT}+e^{(\epsilon_s+U_C+eV)/k_BT}+
e^{2eV/k_BT}}\, .
$$
In this case, the energy conservation law  is independent of $eV$, so
the transitions are suppressed until the voltage reaches
$|\epsilon_s|+U_C$, then it rapidly rises, before $P_{00}-P_{11}$
saturates at unity.  For the pairs outside the depletion zone, we can
write $\mu=0$ and the expression for $P_{00}-P_{11}$ is
\[ \frac{e^{2eV/T}-1}{1+ e^{-\epsilon_s/T} +
e^{(\epsilon_s+2eV+U_C)/T} + e^{2eV/T}},  .
\] While these expressions are somewhat similar, the main difference
lies in the $eV$-dependence of the conservation law which means that
as the voltage changes, the choice of pairs satisfying energy
conservation will change, so a single pair will pass into and out of
the allowed range instead of saturating.

In the following we assume the Fermi level in a semiconductor to
located inside the impurity band, which is sufficiently larger than
$\Delta \gg eV$. Therefore when performing the summation over all
pairs, we choose a uniform distribution of $\epsilon_s$ .  In this
case, both expressions will give exactly the same  results, although
the physics behind them are slightly different.  Neglecting the
Coulomb interaction between the components of the pair, we simply get
a linear relation in both cases.  Thus Ohmic behavior would persist
even though $eV$ can exceed the temperature $T$.  For pairs inside the
depletion zone, this is due to inclusion and saturation of more pairs
as $eV$ becomes larger than the energy splitting of the pairs. If a
given pair has started contributing, one channel has been opened, it
will not terminate  with increasing $eV$. Outside the depletion zone,
pairs will only contribute for the width of the $\delta$-function, so
as voltage rises, other pairs will take over the transport, but the
number of pairs that can contribute will rise linearly with the
voltage.

\begin{figure}[h]
\centerline{
\includegraphics[width=8cm]{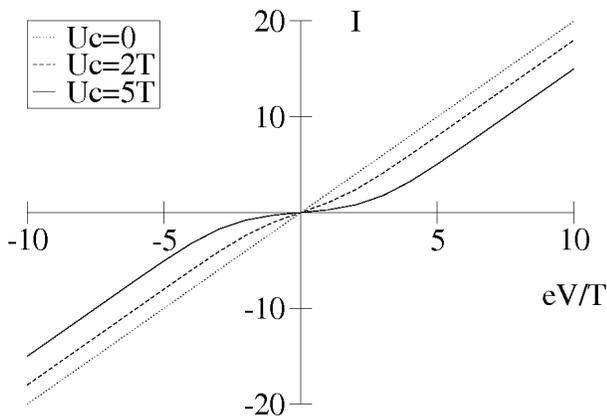}
}
\caption{Current-voltage curve for  $U_C/T = 0$, 1, and
  2. \label{fig:ivcurve}}  
 \end{figure}

A typical $I-V$-curve is shown in Fig.~\ref{fig:ivcurve}. This curve
is calculated under a simplifying assumption that the Coulomb
correlation energy $U_C$ is kept constant of the same order as the
temperature $T$ since the contribution of the $sl$ pair is  $\propto
\rho_{sl}^{-2}e^{-\rho_{sl}/\xi}$ and cut-off at small $\rho$ by the
requirements $U_C \sim T$.~\cite{Kozub1} For this pairs the Coulomb
interaction is essentially screened and does not block two-electron
tunneling.  This assumption significantly simplified the calculation
comparing with averaging over all correlation energies, but does not
change the conclusion. As we see, the transport is suppressed at $eV 
\lesssim U_C$,while for $eV > U_C$ transport will return to a
Ohmic behavior, as shown in Fig.~\ref{fig:ivcurve}.
Thus we predict a rather unusual situation when the conductance
evolves for a low-field Ohmic to a high-field Ohmic regime through an
intermediate non-Ohmic one. Such behavior is a hallmark of the Coulomb
correlation between the occupation numbers of the pairs responsible
for crossed Andreev reflections.
  
\section{Conclusions}


In conclusion, we have shown that the pair tunneling through a barrier
at the interface between a superconductor and hopping insulator is
extremely sensitive to the properties of the tunneling barrier. This
sensitivity is due to rapid oscillations (at scale $\sim k_F^{-1}$) of
the electron wave functions in a superconductor comparing to the
characteristic scale, $a$, of variation of the localized wave function
in a hopping insulator. These oscillations cause dramatic
cancellations in the tunneling probability if the barrier is
uniform. However, this cancellation of the interference origin can be
suppressed if the barrier is inhomogeneous, as it was demonstrated for
different models of a barrier.  In particular, the fluctuations in the
barrier strength  of the scale $\lambda$ falling within the window
$1/k_F \ll \lambda < a$ suppress the cancellations and restore the
transport even if their relative amplitude $\eta$ is very small. For a
barrier with a block-like disorder with the scale $\sim a$ we obtained
an estimate for suppression of the  oscillations $\eta \gtrsim
e^{-k_Fa}$ for the pairs located at the distance $\sim a$ from the
interface.  This happens if the beneficial barrier fluctuations must
coincide with particularly  positioned impurities with the right
energies, means that the number of impurities contributing to
conductance will be relatively small and the relative importance of
``successful'' pairs will increase.  Consequently, one can expect
pronounced mesoscopic - sample specific and reproducible -
fluctuations in both Ohmic and non-Ohmic conductance. Such
fluctuations will be especially pronounced when the barrier contains
large transparency fluctuations (punctures). One can expect that the
fluctuations will have different behavior depending on the location of
the relevant pairs with respect to the position of the depletion zone
near the interface.  We plan to study mesoscopic fluctuations of the
Andreev transport between a superconductor and a hopping insulator in
more detail as a separate project.

Another specific feature of the transport is sensitivity of the
non-Ohmic transport to Coulomb correlation in the occupation numbers
of the relevant pairs. This correlation leads to non-Ohmic behavior at
\textit{low} voltages, $eV \lesssim U_C, T$, while at higher voltages
the transport turns out to be Ohmic. This re-entrant behavior is a
hallmark of the Coulomb correlation.

\acknowledgments
This work was partly supported by the U. S. Department of Energy Office of
Science through contract No. W-31-109-ENG-38 and by the Norwegian Research
Council via a StorFosk program. We are thankful to V. I. Kozub,
V. Vinokur, and A. A. Zyuzin for discussions.

\end{document}